\documentclass{article} 
\usepackage{iclr2026_conference,times}

\usepackage{amsmath,amsfonts,bm}

\def\eqref#1{equation~\ref{#1}}

\def\1{\bm{1}}

\DeclareMathAlphabet{\mathsfit}{\encodingdefault}{\sfdefault}{m}{sl}
\SetMathAlphabet{\mathsfit}{bold}{\encodingdefault}{\sfdefault}{bx}{n}

\usepackage{hyperref}
\usepackage{url}
\usepackage{float}
\usepackage{subcaption}

\usepackage[most]{tcolorbox}
\usepackage{xcolor}
\usepackage{multirow}
\usepackage{bm}
\usepackage{graphicx}
\usepackage{subcaption} 
\usepackage{booktabs}

\newtcolorbox{promptbox}[1][]{
  colback=gray!5!white,    
  colframe=gray!75!black,  
  title={\textbf{Example}}, 
  fonttitle=\bfseries,
  sharp corners,
  boxrule=0.5pt,
  #1
}

\title{Talking with the Latents - how to convert your LLM into an astronomer}

\author{Ilay Kamai  \\
Department of Physics \\
Technion - Israel Institute of Technology\\
Haifa, 32000, Israel \\
\texttt{ilay.kamai@campus.technion.ac.il} \\
\And
Marc-Huertas Company \\
Instituto de Astrofísica de Canarias \\
E-38205 La Laguna, Tenerife, Spain \\
\AND
Michael J. Smith \\
Center for Astrophysics, Harvard \& Smithsonian \\
60 Garden St. Cambridge, MA 02138, USA \\
\And
Hagai B. Perets \\
Technion - Israel Institute of Technology\\
Haifa, 32000, Israel \\
}

\iclrfinalcopy 
\begin{document}

\maketitle

\begin{abstract}
Recent advances in Large Language Models (LLMs) offer unique opportunities for scientific tasks, yet their ability to reason over complex numerical data remains largely unexplored. We propose a simple mechanism to introduce domain-specific physical knowledge into LLMs by fusing pre-trained latent physical features with a pre-trained language model.

Our method employs a teacher-student knowledge distillation framework where a large LLM (teacher) generates synthetic question-answer supervision to transfer physical reasoning to a smaller LLM (student). The student is conditioned on latent physical features and trained via a lightweight adapter and Low-Rank Adaptation (LoRA).
We demonstrate that this approach, applied to models with 1B, 8B, and 32B parameters, enables effective reasoning over real scientific data. Our models substantially outperform strong baselines, such as Gemini 3 Pro, across multiple downstream tasks without task-specific fine-tuning. We show that the model combines latent information with general physical understanding to predict complex properties and can be "steered" by identifying physically meaningful directions in the latent space. This allows for explicit physical manipulation and natural language interpretation of latent structures.
While our experiments focus on astrophysics, the framework is domain-agnostic and applicable to various scientific fields. Our main contribution is a general framework for using LLMs as interpretable interfaces to scientific latent spaces, enabling a single model to perform diverse tasks through natural language guidance. This work marks a step toward developing scientifically capable and useful LLMs.
\end{abstract}

\section{Introduction}\label{sec:intro}

The integration of Large Language Models (LLMs) into the scientific workflow marks a paradigm shift in how researchers interface with information. While these models have demonstrated remarkable proficiency in some scientific tasks, their utility remains largely confined to the textual and visual domains. A critical frontier in "AI for Science" is the transition from processing scientific text to reasoning directly over the numerical and physical data that defines scientific discovery. Despite their sophistication, standard LLM architectures often struggle with numerical reasoning \citep{Rahman2025}, creating a gap between natural language and empirical data.

\textbf{Motivation.} The main motivation of this work is the need to combine physics foundation models and LLMs in a scientifically useful manner. Each one of them has unique advantages, and the potential for knowledge alignment between them has far-reaching outcomes. One of them is interpretability: since LLMs understand both natural language and high-dimensional data, they can serve as a `bridge' between humans and machine learning representations, which are usually hard to interpret. Another possible benefit from such a combination is a flexible downstream task: currently, scientific foundation models serve as a backbone for different downstream tasks. However, each task requires a specific downstream head and training. Using an LLMs as a `lingua franca' interface enables a more flexible downstream behavior, driven by natural language, and encapsulated in a single model.

\textbf{Methodology.} In this work, we develop a general framework to combine scientific foundation models and LLMs. The scientific domain we chose to focus on is astrophysics, but nothing is specific to astrophysics in our method, and it can be transferred to other scientific domains with dedicated machine learning models. As this is one of the first steps towards a language-astronomy multimodal model (other than astronomical images), we chose a simple early fusion approach, where we project the astronomical latent features into the token embedding space of the LLM using an MLP, which we refer to as the Adapter Network (AN). Therefore, the full model is a language model with additional effective tokens resulting from the AN. As such, it is straightforward to generate text conditioned on latent features.

\section{Related Work}\label{sec:related_work}

Advances in LLMs have already had a wide-ranging impact on how we do science. LLM agents can now review and write scientific papers \citep{Villaescusa-Navarro2025_denario, Zhuang2025}, and even act as a 'virtual lab' that suggests and implements novel research ideas \citep{swanson_virtual_2025}. However, these agents typically access scientific data indirectly by writing code to an external tool (e.g., ESM, AlphaFold-multimer). This resonates with the known limitations of current LLMs with numerical data \citep{Rahman2025} and the gap in performance between theoretical questions and data analysis \citep{Pinheiro_2025_olympiad_astronomy}.
On the other hand, scientific foundation models become popular in many scientific fields and show strong performance across domains \citep[e.g.,][]{Leung2024, Parker2025_aion, Smith2024_astropt, euclid_2025_multimodal, Bardhan2025, Hayes2025_ESM3, bodnar_2025_aurora, abramson_2024_alphafold3}. Notably however, those models do not use natural language as a modality. As a consequence, there is currently a clear distinction between scientific foundational models and LLMs.
The idea of a scientific-LLM usually refers to the injection of scientific knowledge into existing LLMs using various methods (for review, see \cite{Hu2025_sci_llm} and \cite{Eger2025_sci_llm}). However, most works in the field of scientific-LLM focus on fine-tuning LLM on high-quality textual and visual data to create improved reasoning \citep{Xie2023_darwin, Bai2025_intern_s1, Cherian2024_llmphy}, object descriptions \citep{Mishra-Sharma2024}, search engines \citep{Koblischke2025}, and scientific chatbots \citep{Zaman2025}. While such models often achieve impressive results compared to general LLMs \citep{haan2025_astrosage}, they are usually focused on two specific modalities---text and images. While those modalities are very useful for different scientific tasks, real scientific research has a much larger modality landscape, and the combination of such modalities with LLMs for real scientific tasks is still largely unexplored.\\
Our approach differs from the above by directly aligning the latent space of a scientific model with the LLM. This fusion is motivated by methods that project latent features into token space to improve interpretability \citep[e.g.,][]{Tennenholtz2023}, treating the output of the scientific encoder as effective tokens for the language model.

\section{Question-Answer dataset}\label{sec:dataset}
In this section, we describe the dataset used in our experiments. Our dataset consists of astronomical latent features that are the output of a stellar spectra foundation model, and a textual description of the same samples. In this work, we use two different spectra foundation models. The first model is a pretrained spectra encoder from \cite{Kamai2025_desa}. In their paper, the authors presented a multimodal model for stellar spectra and light curves, which consists of two separate encoders, a spectra encoder and a light curve encoder, and an alignment module. For this work, we utilized the spectra encoder, which comprises a CNN encoder-decoder enhanced with a Conformer module \citep{Gulati2020} that processes the averaged bottleneck from the CNN encoder and yields stellar predictions. We therefore refer to this model as \emph{Spectral Conformer (SC)}. The SC model was trained on approximately $5$ million spectroscopic measurements from the LAMOST survey \citep{Zhao2012_LAMOST}, with two objectives---a self-supervised masked reconstruction and a supervised prediction of stellar parameters---temperature (\emph{Teff}), surface gravity (\emph{logg}), and metallicity (\emph{FeH}). The second model, which we use in section \ref{subsec:res_latent}, use the same input but with a slightly different architecture - it first divides the masked spectra into patches, using a 1D-CNN, and than process it with a transformer encoder. The resulting latent tokens are sent into an MLP decoder, which produces clean spectra, and an MLP regressor that predicts stellar parameters. We refer to this model as \emph{Spectral ViT (SViT)}. The SViT model was trained on the same dataset and with the same objectives. The resulting features, from both SC and SViT models, are of dimension $2048$. To construct the textual description, we used the stellar parameters, as given by the LAMOST stellar parameter pipeline (LASP) \citep{Wu2014_LASP}, and generated a short question-answer pair for this specific star. The question-answer pair was generated by prompting Gemini 2.5 flash with the stellar information and asking it to describe the star as a professional astronomer with no more than $50$ words. In addition, to enable the fusion of latent features into the input, we added $K$ empty tokens at the beginning of the question that were used as placeholders for the latent features. As a compromise between performance and compute, we chose $K=8$. Finally, to further improve the diversity of the dataset, we randomly add a follow-up question for $70\%$ of the samples about the stellar type of the star. The follow-up question includes all the previous context, and the answer to the follow-up question includes only the stellar type without further description. The total size of the dataset consisted of 40,000 samples, which were split into training ($80\%$), validation ($10\%$), and test datasets ($10\%$). An example of a dataset sample is shown in Figure \ref{fig:dataset_sample}.

\begin{figure}[htbp]
    \centering
    \begin{promptbox}[]
        \textbf{Main Question:} \\
        Describe this star based on its stellar parameters.
        
        \tcbline
        
        \textbf{Main Answer:} \\
        This star has a $T_{\text{eff}}$ of $6463$\,K, $\log g$ of $4.04$, and $[\text{Fe/H}]$ of $0.07$. It is likely a main-sequence star similar to the Sun, but slightly hotter and more metal-rich.
        
        \tcbline
        
        \textbf{Follow-up Question:} \\
        What stellar classification best describes this object? \\

        \tcbline
        \textbf{Follow-up Answer:} \\
        F-type dwarf
    \end{promptbox}
    \caption{An example of our question-answer dataset. The follow-up questions are added randomly in $70\%$ of the cases.}
    \label{fig:dataset_sample}
\end{figure}

\section{Latent Interpeter model}\label{sec:model}
\begin{figure}[htbp]
    \centering
    \includegraphics[width=0.45\linewidth]{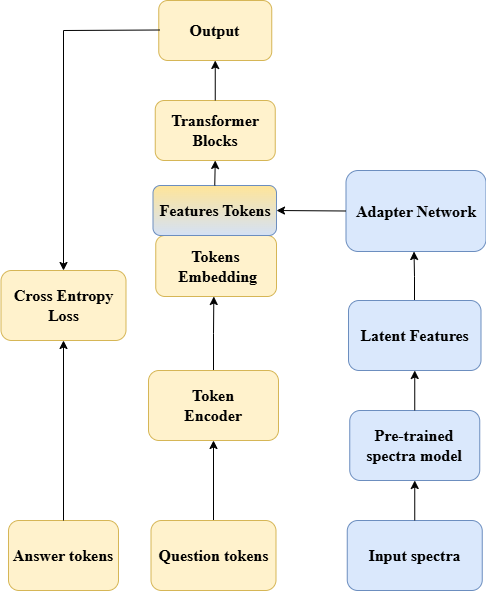}
    \caption{High-level diagram of the Latent Interpeter model. Blue colors represent the spectra model. Yellow colors represent the LLM.}
    \label{fig:model_architecture}\label{fig:model}
\end{figure}

Our model, which we call \emph{Latent Interpeter} (LI), consists of a simple composition of a pre-trained LLM and a pre-trained spectra model. The input of the LLM is the question tokens as described in section \ref{sec:dataset}. They are embedded in a high-dimensional space using the existing token encoder of the LLM. The input of the spectra model is the corresponding spectra. Masked spectra are processed by the spectra model (SC or SViT), resulting in latent features. Features are further transformed into effective token embeddings using a dedicated \emph{Adapter Network} (AN). The AN projects the latent features into $K$ tokens with the same dimensions as the token embedding space. The resulting $K$ effective tokens are concatenated to the original token embeddings and sent into the next LLM layers.  This simple fusion is motivated by \cite{Tennenholtz2023}, which suggested a similar approach to improve latent space interpretability for LLMs. Training is done in two steps---first, only the AN is trained for several epochs. Next, the AN is frozen, and we finetune the LLM with a low-rank adaptation (LoRA)  \citep{Hu2021_lora}. This strict distinction between training the AN and finetuning the LLM is crucial since a good transformation of the latent features into the LLM embedding needs to be learned before we can effectively fine-tune the LLM. A diagram of the model is shown in Figure \ref{fig:model}.

\section{Experimental Results}

The following subsections discuss our general findings.

\subsection{Effect of LLM size} \label{subsec:res_llm}
\textbf{Learning local concepts.}
First, we want to test the ability to infer correct physical information. For that, we train three models, each one using a different pre-trained LLM---Llama-1B, Llama-8B, and Qwen-32B. For the latent features, we used the SC model. The latent features in this case were taken as the output of the Conformer block. All the models were trained with an identical dataset and hyperparameters. After training, we test the models on a separate test set, extract the values of the stellar parameters (\emph{Teff, logg, FeH}) from the generated text, and compare them with the ground truth values. Besides being the most basic stellar parameters to infer from spectra, those parameters were also part of the original training of the SC model. Therefore, we expect strong alignment between latent features and the stellar parameters values, and as a consequence, we call them \emph{local concepts} as they are represented entirely by the latent space. Using this definition, the first experiment tests the ability of the LLM to extract the information stored in specific latent features using the AN. To compare our method with baselines, we use two additional models. First, we use the results of the SC model from \cite{Kamai2025_desa}. We refer to this model as our \emph{spectra baseline model}. Second, we use the Gemini 3 pro model (preview release). To test its ability to infer stellar parameters, we provide it with the following information: arrays of the measured wavelengths and flux, and an image that shows a plot of the wavelength vs. flux. In addition, we provide it with a general context in the form of short guidelines for stellar spectra analysis, which was created using Gemini 3 pro's deep research feature (approximately $40\,000$ tokens). The prompt asks the model to act as a professional astrophysicist and to analyze the spectrum using all the given information. We refer to this model as \emph{Gemini baseline model}. 

Table \ref{tab:basic_stellar_params} compares the performance between our three trained models, tested on $100$ samples, the spectra baseline model, and the Gemini baseline model, tested on $50$ samples, for efficiency. We measure the median absolute error (MedAE), and the root mean squared error (RMSE), which is more sensitive to outliers. We see that the Gemini baseline model performs much worse than all the trained models. In addition, our trained models achieve comparable or better results compared to those of the spectra baseline model, with no one model performing consistently better than the others. The immediate conclusion is that all three models were able to effectively extract information from local concepts stored in the latent features, much better than a state-of-the-art LLM that uses general knowledge and a multimodal input (Gemini baseline model). The comparison to a state-of-the-art (SOTA) LLM is, of course, not fair from a data perspective---the SOTA LLM didn't see the specific latent features but only the raw data. However, the purpose of such a comparison is to convince that for scientific tasks related to specific data, general pre-training is not sufficient, even for SOTA LLMs, and that our special training approach is required.

\begin{table}[h]
    \centering
     \caption{Performance comparison for $T_{eff}$, $\log g$ and metallicity (Fe/H). All values were extracted from the models' answers.}
    \label{tab:basic_stellar_params}
    \resizebox{\textwidth}{!}{%
        \begin{tabular}{lcccccc} 
            \toprule 
            \multirow{2}{*}{\textbf{Model}} & 
            \multicolumn{2}{c}{\textbf{\bm{$T_{\text{eff}}$} [K]}} & 
            \multicolumn{2}{c}{\textbf{\bm{$\log g$} [dex]}} & 
            \multicolumn{2}{c}{\textbf{[Fe/H] [dex]}} \\
            
            \cmidrule(lr){2-3} \cmidrule(lr){4-5} \cmidrule(lr){6-7} 
            
             & MedAE $\downarrow$ & RMSE $\downarrow$ & MedAE $\downarrow$ & RMSE $\downarrow$ & MedAE $\downarrow$ & RMSE $\downarrow$ \\
            \midrule 
            
            LI-Llama-1B & 30.31 & 114.64 & 0.04 & 0.19 & 0.04 & 0.14 \\
            LI-Llama-8B & 26.50 & \textbf{63.96} & 0.04 & \textbf{0.11} & \textbf{0.03} & \textbf{0.06} \\
            LI-Qwen-32B & \textbf{23.05} & 112.07 & \textbf{0.03} & 0.13 & 0.05 & 0.53 \\
            Spectra Baseline model & 35.58 & 85.18 & 0.05 & 0.12 & \textbf{0.03} & \textbf{0.06} \\
            
            \midrule 
            
            Gemini 3 Pro (Zero-Shot) & 357.70 & 564.97 & 0.28 & 0.75 & 0.15 & 0.22 \\
            \bottomrule 
        \end{tabular}
    }
\end{table}

\textbf{Learning global concepts.} Next, we want to test if the models can combine information from latent features with their own general physical knowledge. For that, we perform a similar experiment as used to infer basic stellar parameters, but we also test the performance on three follow-up questions. First, we tested the prediction of the spectral type of the star. In addition to spectral type, which appeared in the training, we also test the performance on two additional follow-up questions that the model was not trained on---we provide the model with the radius of the star and ask it to predict the luminosity and mass of the star. The values of the radius, luminosity, and mass were taken from \cite{Berger2020}. All three tasks require combined information - inferring the spectral type requires a combination of $T_{eff}$ and $\log g$. Calculating mass and luminosity from the radius alone is not possible, and requires knowledge about $T_{eff}$ and $\log g$ (specific knowledge) as well as the analytical relationships between the parameters (general knowledge). Therefore, these questions test whether the models can combine the local concepts with general knowledge that comes from general pre-training. Table \ref{tab:advanced_stellar_params} shows the results. Here, we observe a more consistent trend: the largest model (Qwen-32B) performs best across all tasks, with the exception of the luminosity RMSE. Given that the MedAE of Qwen-32B remains better than other models, we conclude that the higher RMSE is driven primarily by a greater number of outliers for this model. This experiment highlights a capability unique to the LLM architecture: multimodal reasoning. To predict luminosity or mass, the model must fuse information from three distinct sources: the specific spectral data stored in the latent features ($T_{eff}$), the auxiliary data provided via the text prompt (Radius), and the general physical laws stored in the LLM's weights (e.g., Stefan-Boltzmann law). A standard regression head trained on spectral features would fail this task, as it lacks access to the auxiliary text input (Radius), while a regular LLM would fail, as it lacks access to the spectral data. By combining text and numerical features, our model can perform zero-shot inference on these tasks.

\begin{table}[h]
    \centering
    \caption{Performance comparison for luminosity, mass, and spectral type.}
    \label{tab:advanced_stellar_params}
    \begin{tabular}{lccccc}
        \toprule
        \multirow{2}{*}{\textbf{Model}} & 
        \multicolumn{2}{c}{\textbf{\bm{$L$} [$L_{\odot}$]}} & 
        \multicolumn{2}{c}{\textbf{\bm{$M$} [$M_{\odot}$]}} & 
        \multirow{2}{*}{\textbf{Spectral Type Acc.}} \\ 
        
        \cmidrule(lr){2-3} \cmidrule(lr){4-5}
        
         & MedAE $\downarrow$ & RMSE $\downarrow$ & MedAE $\downarrow$ & RMSE $\downarrow$ & \\
        \midrule
        
        LI-Llama-1B & 4.54 & \textbf{62.24} & 0.73 & 6.48 & 17.3\% \\
        LI-Llama-8B & 5.62 & 109.15 & 1.08 & 3.21 & 37.4\% \\
        LI-Qwen-32B & \textbf{2.53} & 70.36 & \textbf{0.38} & \textbf{1.58} & \textbf{47.5\%} \\
        
        \bottomrule
    \end{tabular}
\end{table}

The fact that, for all the global concepts, the best performance was achieved by the largest model (Qwen-32B), implies that increasing the LLM size should improve the flexibility of the final model and enable answering more complex questions without dedicated training.

\subsection{Causal control}\label{subsec:res_steering}
\begin{figure}
    \centering
        \centering
        \includegraphics[width=\textwidth]{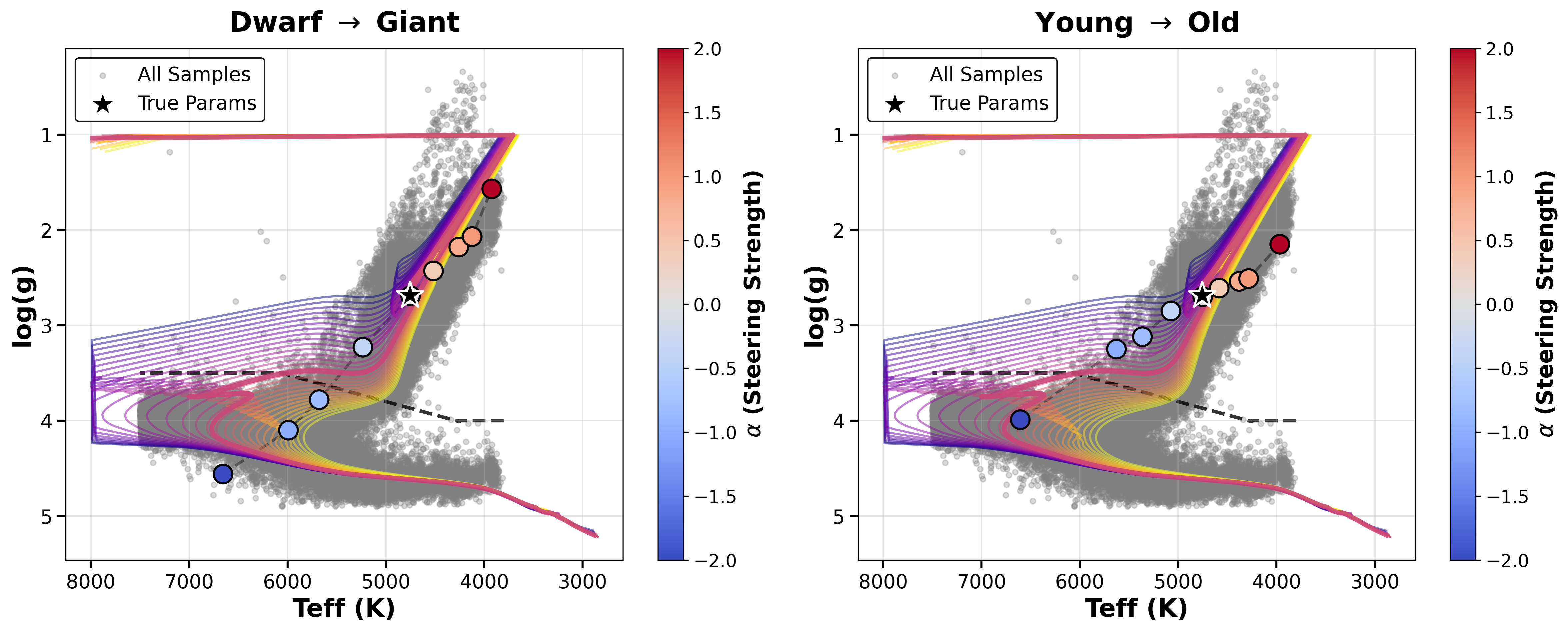}
        \caption{Example of the steering effect for two different concepts - Dwarf to Giants (left) and Young to Old (right). In both panels, the black star represents the true parameters of the star under consideration. The colored circles represent the generated parameters with different $\alpha$ values. The colored lines represent theoretical isochrones - curves with constant age. The dashed black line represents a separation between dwarfs and giants from \cite{Ciardi2011}. The background gray points represent the true parameters of the entire sample set. }\label{fig:steering_example}
\end{figure}

Next, we want to test if we can steer the model's output in a controllable and physical way. To test this, we conduct a steering experiment. Here, we are largely following the procedure suggested in \cite{Fear2025_steering}---first, we identify `concept directions' in the features' latent space of the SC model. To identify such a direction, we collect samples from the training set that obey a constraint defined by some physical concept (for example, $T_{eff} > 6000$ K for a `hot star'), and average all their latent features to get an average representation of a concept. We then repeat the process with the opposite concept (for example, $T_{eff}< 4250$ K for `cool star'), and we subtract the two averages to get the direction of the concept in latent space. We then use this direction as a steering vector---during inference, we add this direction to the original latent features. Specifically, given a latent feature, $f$, and a direction feature $f_{direc}$, the final features given to the model are $f' = f + \alpha \frac{f_{direc}}{\| f_{direc}\|^2}$, where $\alpha$ controls the steering strength. For each sample, we tested different concepts and changed $\alpha$ between $-2$ and $2$. To interpret the results, we extract the physical parameters from the generated answer and plot them in $T_{eff}-\log g$ space. Figure \ref{fig:steering_example} shows an example of steering with two different concepts---stellar evolution concept (dwarfs to giants) and stellar age concept (young to old). The evolution concept was derived from \cite{Ciardi2011}, which presented a criterion to identify giants based on $T_{eff}$ and $\log g$ cuts. We took all the samples that matched the criterion as giants and all others as dwarfs. This is, of course, a crude estimation as the evolution of stars is continuous and includes a `sub-giant' regime. For the age concept, we used isochrone ages from \cite{Berger2020}. We took samples with ages younger than the $25^{th}$ percentile as `young' and samples with ages older than the $75^{th}$ percentile as `old'. This is again a rough estimation, as isochrone ages are accurate only at certain evolutionary phases (main sequence turn off, for example). In other phases, isochrones are densely packed, and the derived age can have very large errors. Since we focused on the ability to steer the model, and not on the effect of LLM size, in this experiment, we used only Llama-8B LLM. 
We can see that the effect of steering matches our physical understanding---when the model is steered towards a dwarf or a giant (left panel), it moves in the right direction, almost perpendicular to the black dashed line. When it steered with age direction, the model again moves in the expected direction (bluer circles move toward magenta isochrones), this time more perpendicular to the isochrones, as expected. In both cases, the full steering paths reside in a physical region, where the population of real stars is large. One example of that is in the right panel; when $\alpha=-1$, the parameters sit on the edge of the gray background, and we see that for the next point, $\alpha=-2$, we have a big `jump' in parameters, much bigger compared to previous steps, to avoid the non-physical area. We observed that our model produced more physical trajectories compared to an MLP that was trained to predict stellar parameters from spectral features, and was steered in the same way (see Appendix \ref{subsec:app_figs}). It is also important to emphasize that steering changes the entire generated answer and not only the extracted parameters. For example, when the model is steered towards a dwarf star, the entire description would be of a dwarf (see, for example, Figure \ref{fig:steering_text}). 
We tested additional concepts --- hot to cool, low $\log g$ to high $\log g$, metal poor to metal rich, and low mass to high mass. In all cases, we see the same physically-consistent behavior (see Appendix \ref{subsec:app_figs}).

\begin{figure}[htbp]
    \centering
    \begin{promptbox}[]
        \textbf{Main Question:} \\
        Describe this star based on its given stellar parameters. \\

        \textbf{True stellar parameters:}\\
        Teff : $4757$,
        logg : $2.69$, 
         FeH : $0.06$

        \tcbline
        
        \textbf{Model Answer ($\alpha=0$):} \\
        This star has a Teff of $4742$ K, logg of $2.68$, and [Fe/H] of $0.00$. These parameters suggest it is a evolved star, likely a red giant. The near-solar metallicity indicates it formed from material similar to the Sun. \\
        
        \textbf{Model Answer ($\alpha=-2$):} \\
        This star has an effective temperature (Teff) of $6672$ K, a surface gravity (logg) of $4.54$, and a metallicity ([Fe/H]) of $0.10$. These parameters suggest a main-sequence star similar to our Sun. The high metallicity indicates a relatively old age. The given temperature is slightly hotter than the Sun's. The surface gravity is also higher, suggesting a more massive star. These parameters suggest a star similar to our Sun, but slightly older and more massive. The high metallicity suggests a relatively old age. The given temperature is slightly hotter than the Sun's. \\

        \textbf{Model Answer ($\alpha=2$):} \\
        This is a red giant star with Teff = $3912$ K, log g = $2.00$, and [Fe/H] = $-0.06$. The low surface gravity indicates it has evolved off the main sequence. The near-solar metallicity suggests it is a member of the galactic disk.
        
        \tcbline
      
    \end{promptbox}
    \caption{Answers generated by the model during the steering experiment, for the example in Figure \ref{fig:steering_example}, left panel, for $\alpha=0,-2,2$}
    \label{fig:steering_text}
\end{figure}

\subsection{Effect of latent dimension}\label{subsec:res_latent}
\begin{figure}[h]
    \centering
    \begin{subfigure}[b]{0.49\textwidth}
        \centering
        \includegraphics[width=\linewidth]{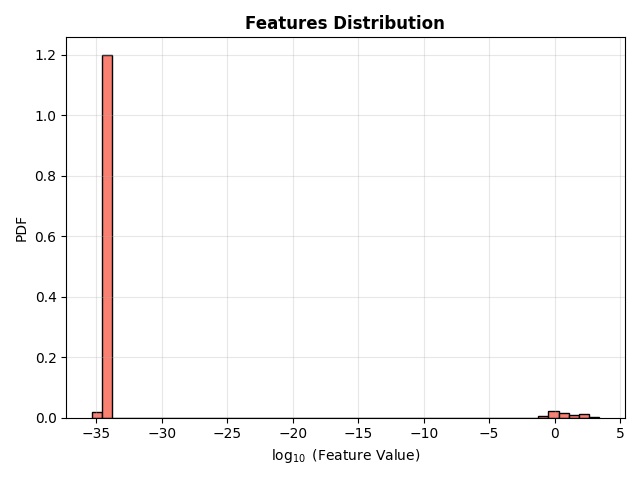}
        \caption{SC Features} 
    \end{subfigure}
    \hfill
    \begin{subfigure}[b]{0.49\textwidth}
        \centering
        \includegraphics[width=\linewidth]{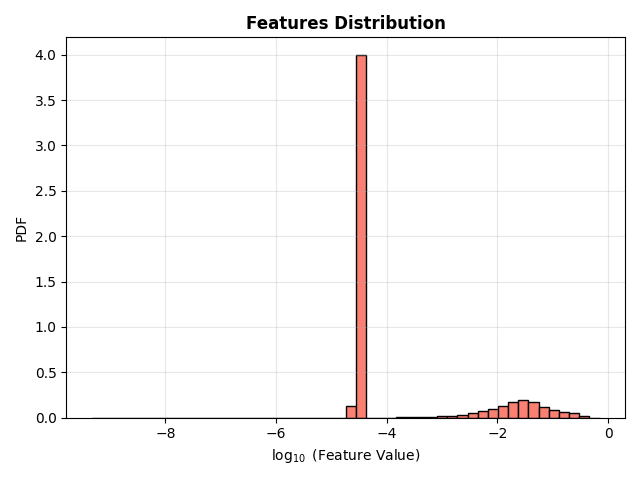}
        \caption{SViT Features} 
    \end{subfigure}
    
    \caption{Distributions of latent feature values for SC (left) and SViT  (right) models.}
    \label{fig:hist_features}
\end{figure}

In section \ref{subsec:res_llm}, we tested the effect of changing the LLM size with fixed latent space. Now we investigate the opposite direction and ask how the structure and characteristics of the latent space affect the final results, given a fixed-size LLM.
To test this, we used the SViT model as described in section \ref{sec:dataset}. The final features of the SViT model were taken to be the classification token that was used to predict stellar parameters. While the features from the SC model are sparse (approximately $60\%$ of the elements are $<10^{-30}$), the features from the SViT model are dense. Figure \ref{fig:hist_features} shows a statistical comparison between the sparse and dense features. An important point to emphasize is that both features have the same dimensionality ($2048$), and that both networks show similar performance on stellar parameters prediction.\\
We trained two models with a Llama-8B LLM, each one with different latent features. Training of both models was identical, except for the latent features given to the LLM. Table \ref{tab:basic_stellar_params_sparsity} shows the results of basic parameter prediction (local concepts). We see a huge gap in performance between the different features. Using the same intuition as in section \ref{subsec:res_llm} that the basic features represent specific knowledge which gained only during our training, this striking difference between the models implies that sparsity is crucial for the successful interpretation of features as effective tokens. While intuitive, this conclusion should be take with caution as the features come from two different models, and might have other subtle differences. Therefore, to conclusively attribute the performance differences to sparsity, additional experiments are needed. However, the striking performance gap, which was not observed in MLP regression during SC and SViT training, suggests that architectural choices in the feature encoder significantly affect the LLM's ability to interpret latent representations. 

\begin{table}[h]
    \centering
     \caption{Performance comparison for $T_{eff}$, $\log g$ and metallicity (Fe/H). All values were extracted from the models' answers.}
    \label{tab:basic_stellar_params_sparsity}
    \resizebox{\textwidth}{!}{%
        \begin{tabular}{lcccccc}
            \toprule
            \multirow{2}{*}{\textbf{Model}} & 
            \multicolumn{2}{c}{\textbf{\bm{$T_{\text{eff}}$} [K]}} & 
            \multicolumn{2}{c}{\textbf{\bm{$\log g$} [dex]}} & 
            \multicolumn{2}{c}{\textbf{[Fe/H] [dex]}} \\
            
            \cmidrule(lr){2-3} \cmidrule(lr){4-5} \cmidrule(lr){6-7} 
            
             & MAE $\downarrow$ & RMSE $\downarrow$ & MAE $\downarrow$ & RMSE $\downarrow$ & MAE $\downarrow$ & RMSE $\downarrow$ \\
            \midrule
            
            LI-Llama-8B-SC & \textbf{26.50} & \textbf{63.96} & \textbf{0.04} & \textbf{0.11} & \textbf{0.03} & \textbf{0.06} \\
            LI-Llama-8B-SViT & 161.92 & 638.44 & 0.38 & 0.76 & 0.25 & 0.42 \\
            
            \bottomrule
        \end{tabular}
    }

\end{table}

\subsection{Interpretability of latent space}
\begin{figure}
    \centering
    \begin{minipage}[b]{\textwidth}
        \centering
        \includegraphics[width=0.8\textwidth]{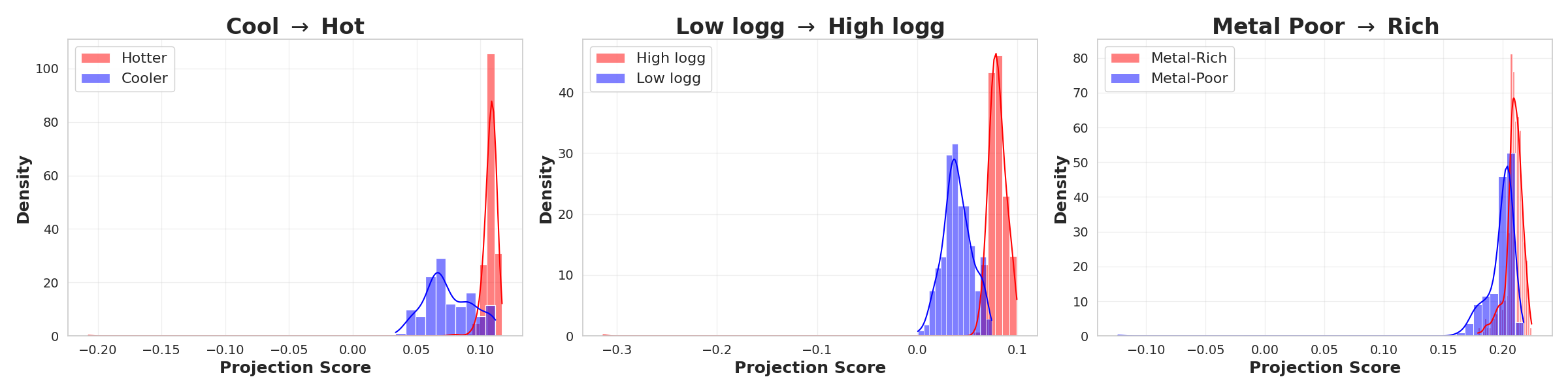}
    \end{minipage}
    \hfill
    \begin{minipage}[b]{\textwidth}
        \centering
        \includegraphics[width=0.8\textwidth]{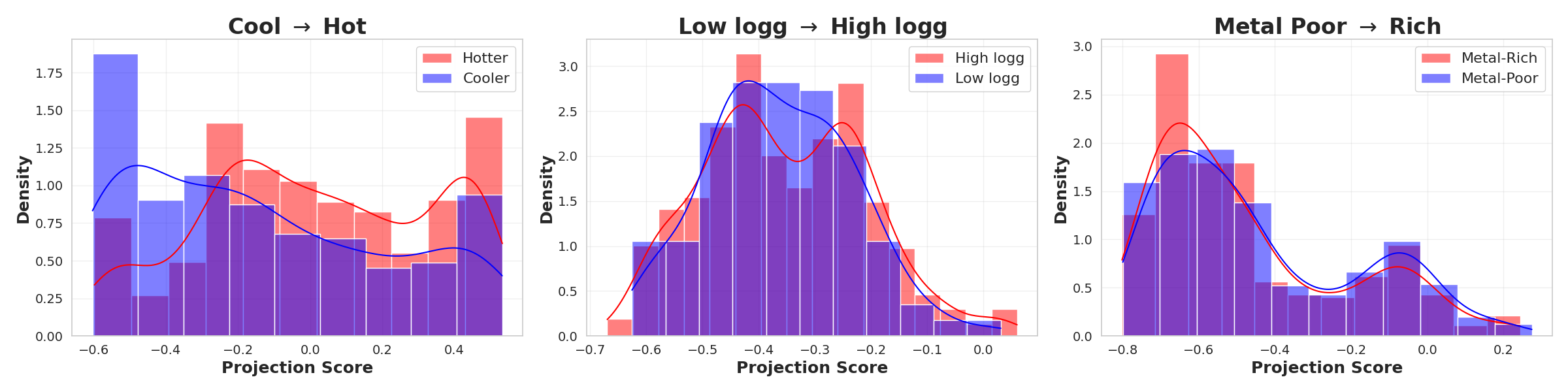}
    \end{minipage}
    \caption{Projections of latent features from the test set on concepts from the training set. The upper row corresponds to SC features, and the lower row corresponds to SViT features.}
    \label{fig:concepts_projection_hist}
\end{figure}

The previous experiment is encouraging, as it highlights a subtle difference between latent features that is not easily identified with standard tools. Indeed, both spectra models achieved comparable performance during their training. Yet, the difference in the ability of the LLM to interpret the features implies a significant difference between them. This difference can be seen when looking at concept directions, like in section \ref{subsec:res_steering}. To test the characteristic of the concept direction, we project features from the test set on the direction vectors. We then classify the projections according to the same physical criteria that were used to create the concept and compare the populations. This is a way to measure whether opposite concepts (for example `hot star' vs `cool star') have distinct features. Figure \ref{fig:concepts_projection_hist} shows that the sparse features (from the SC model) results in distinct separation between concepts, while the dense features (from the SViT model) result in more mixed concepts. This difference can be qualitatively measured using the Kolmogorov-Smirnov test, which tests the hypothesis that two populations are drawn from the same distribution. Table \ref{table:ks_test_concepts_projection} shows that in the sparse feature space, opposite concepts present two clearly separated distributions (p-value $< 10^{-85}$), and in the dense feature space, the separation is not always significant (p-value $\sim 10^{-1}$ for $logg$ and $FeH$). This analysis suggests that the concepts in the sparse feature space are likely more linearly separable compared to the dense alternative. Our model naturally highlighted this fact with the poor performance of the LLM using dense features. This suggests that our model can serve as an effective \textit{latent interpreter}, and that it is possible to probe the latent space using LLM. Of course, interpretability using performance analysis is not ideal, and more effective methods should be investigated. However, the fact that we see such a performance gap implies that such methods are possible.       

\begin{table}[h]
    \centering
    \caption{$\log_{10}$ of the p-value of a Kolmogorov-Smirnov test for the hypothesis that projections of features on two distinct concepts are drawn from the same distribution. Lower values mean a lower probability that this hypothesis is true.}
    \label{table:ks_test_concepts_projection}
    \begin{tabular}{lccc}
        \toprule
        \textbf{Model} & $\log_{10}(\text{p-value}) - T_{\text{eff}}$ & $\log_{10}(\text{p-value}) - \log g$ & $\log_{10}(\text{p-value}) - \text{FeH}$ \\
        \midrule
        LI-Llama-8B-Sparse & -135.2 & -154.4 & -85.6 \\
        LI-Llama-8B-Dense  & -10.3  & -0.86  & -1.1  \\
        \bottomrule
    \end{tabular}
\end{table}

\section{Discussion}\label{sec:discussion}
In this work, we presented \emph{Latent Interpeter}---a framework for aligning scientific foundation models with large language models, using stellar astrophysics as a case study. Our results demonstrate that a lightweight adapter network, combined with low-rank adaptation (LoRA), can effectively translate high-dimensional physical latent spaces into the semantic space of an LLM.\\
Our main results are as follows: first, we show that by combining latent features information, general physics understanding of the LLM, and the ability to provide the LLM additional information easily, our model can solve complex scientific tasks (e.g., predict the mass of the star given its radius), that are not feasible for spectra-only or text-only models. We also show that this ability scales with LLM size. Second, the steering experiments in Section \ref{subsec:res_steering} provide compelling evidence that the model learns the \emph{topology} of the physical parameter space, not just statistical correlations. The ability to smoothly traverse the latent space along physically meaningful axes suggests that the adapter preserves the continuous nature of physical laws when mapping to the discrete token space. Lastly, we showed that using two different scientific features with different topologies (sparse vs. dense features) results in a significant performance gap of the LLM, which demonstrates the ability of using LLMs as sophisticated interpretability machines for scientific latent space.\\
\textbf{Limitations and Future Work.}
This paper serves as a proof of concept. As such, relatively small LLMs and a simple unimodal spectra model were used. To fully utilize the potential of this approach, larger LLMs and scientific foundation models should be used. In addition, our approach relies on a `Teacher-Student' distillation pipeline. While effective, the student model is fundamentally bounded by the quality and biases of the teacher's synthetic descriptions. Furthermore, our current architecture is unidirectional: it can describe data but cannot generate scientific data from text. Enabling such a bidirectional flow requires a different architecture, which is a focus of future work.

\clearpage

\bibliography{main}
\bibliographystyle{iclr2026_conference}

\appendix
\section{Appendix} \label{sec:appendix}

\subsection{implementation details} \label{subsec:app_implementation}
The raw spectra were taken from LAMOST DR8 \citep{Zhao2012_LAMOST, Wang2022_lamost_dr8}. In all models, we used $8$ effective tokens for the features. For the Adapter Network, we used a 2-layer MLP with a hidden dimension of $512$ and a GELU activation function \citep{Hendrycks_2016_gelu}. For LoRA fine-tuning, we used a rank of $8$ and $\alpha=8$. We used an AdamW optimizer \citep{Loshchilov2017_adamw} with a learning rate of $10^{-4}$ and a scheduler that performs a linear warmup followed by cosine decay. Training was done in two steps - we first trained only the AN, and then applied LoRA with the AN frozen. We set the number of epochs in the first step to $5$. The convergence of LoRA typically occurs within several additional epochs. All models were trained using 4 L40s or 4 H200 GPUs for a maximum of $30$ epochs.  

\subsection{concepts criterion} \label{subsec:app_concepts}

\begin{table}[H]
    \centering
    \renewcommand{\arraystretch}{1.5} 
    \setlength{\tabcolsep}{10pt}      
    
    \begin{tabular}{|c|c|}
    \hline
        \textbf{Concept} & \textbf{Criteria} 
          \\ 
        \hline
        Cool $\xrightarrow[]{}$ Hot & $T_{\text{eff}} > 6250$ (Hot), $T_{\text{eff}} < 4500$ (Cool) \\
        \hline
        Low $\log g$ $\xrightarrow[]{}$ High $\log g$ & $\log g > 4.5$ (High), $\log g < 3.0$ (Low) \\
        \hline
        Metal Poor $\xrightarrow[]{}$ Metal Rich & $[\text{Fe/H}] > 0.2$ (Rich), $[\text{Fe/H}] < -0.2$ (Poor) \\
        \hline
        Dwarf $\xrightarrow[]{}$ Giant & $\log g < \text{Thresh}(T_{\text{eff}})$ (Giant), $\log g \ge \text{Thresh}(T_{\text{eff}})$ (Dwarf) \\
        \hline
        Low Mass $\xrightarrow[]{}$ High Mass & Top 25\% vs. Bottom 25\% (controlled for Age \& $[\text{Fe/H}]$) \\
        \hline
        Young $\xrightarrow[]{}$ Old & Top 25\% vs. Bottom 25\% (controlled for Mass \& $[\text{Fe/H}]$) \\
        \hline
    \end{tabular}
    \caption{Criteria used to define the concept directions. For the evolution concept, Thresh($T_{eff}$) was taken from \cite{Ciardi2011}. For Mass and Age, data is stratified into local bins of determining parameters (e.g., Age bins of 1 Gyr, Mass bins of 0.1 $M_{\odot}$) to isolate the concept.}
    \label{tab:concept_criteria}
\end{table}

\subsection{Additional Figures} \label{subsec:app_figs}

\begin{figure}[h]
    \centering
    \begin{subfigure}[b]{\textwidth}
        \centering
        \includegraphics[width=\linewidth]{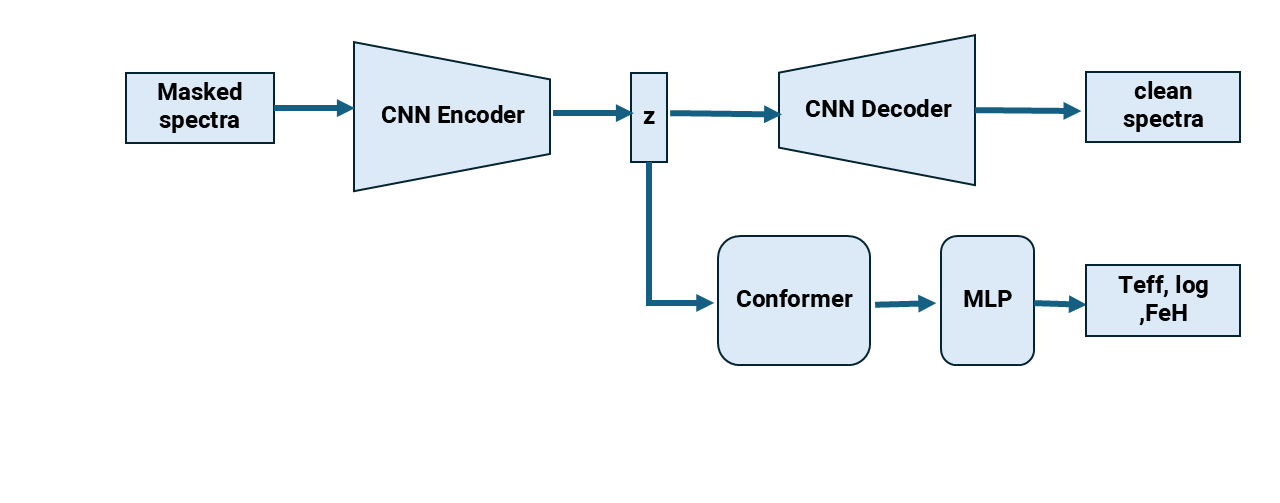}
        \caption{Spectral Conformer (SC)}
        \label{fig:spectra_encoder}
    \end{subfigure}
    \hfill 
    \begin{subfigure}[b]{\textwidth}
        \centering
        \includegraphics[width=\linewidth]{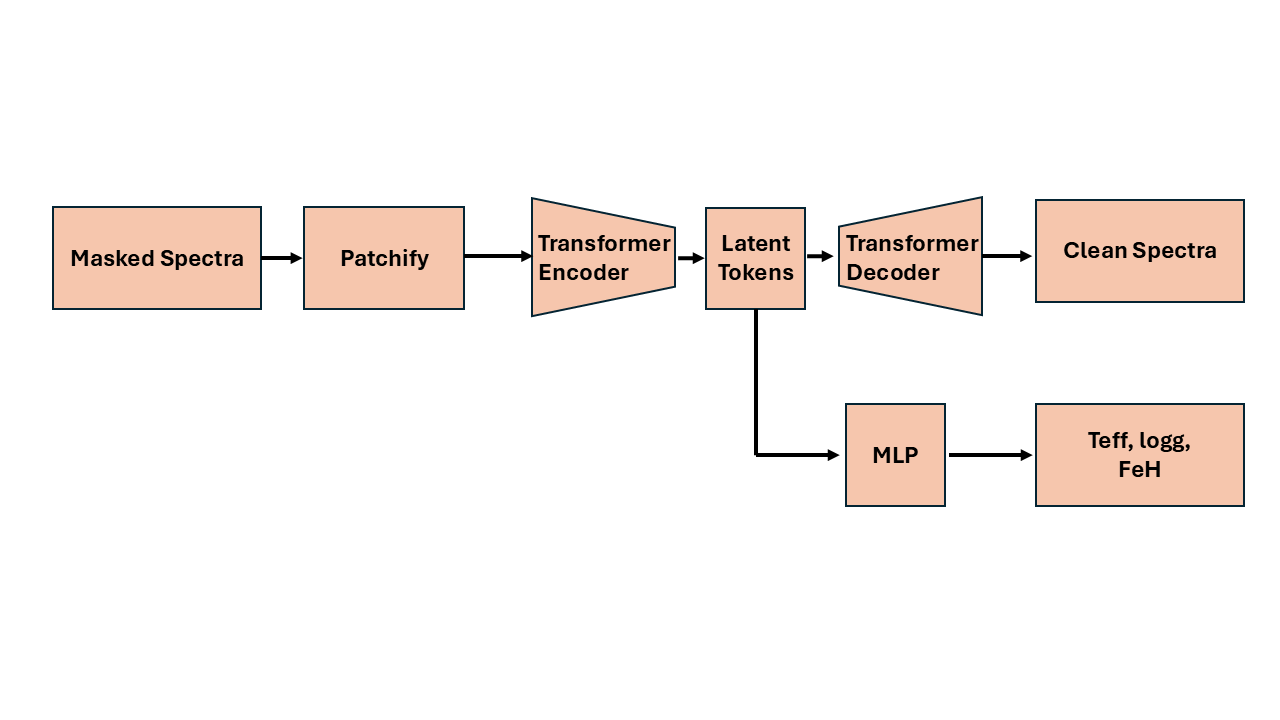}
        \caption{Spectral ViT (SViT)}
        \label{fig:spectral_vit}
    \end{subfigure}
    
    \caption{Diagrams of the spectra models used in this work.}
    \label{fig:spectra_comparison}
\end{figure}

\begin{figure}[h]
    \centering
        \centering
        \includegraphics[width=\textwidth]{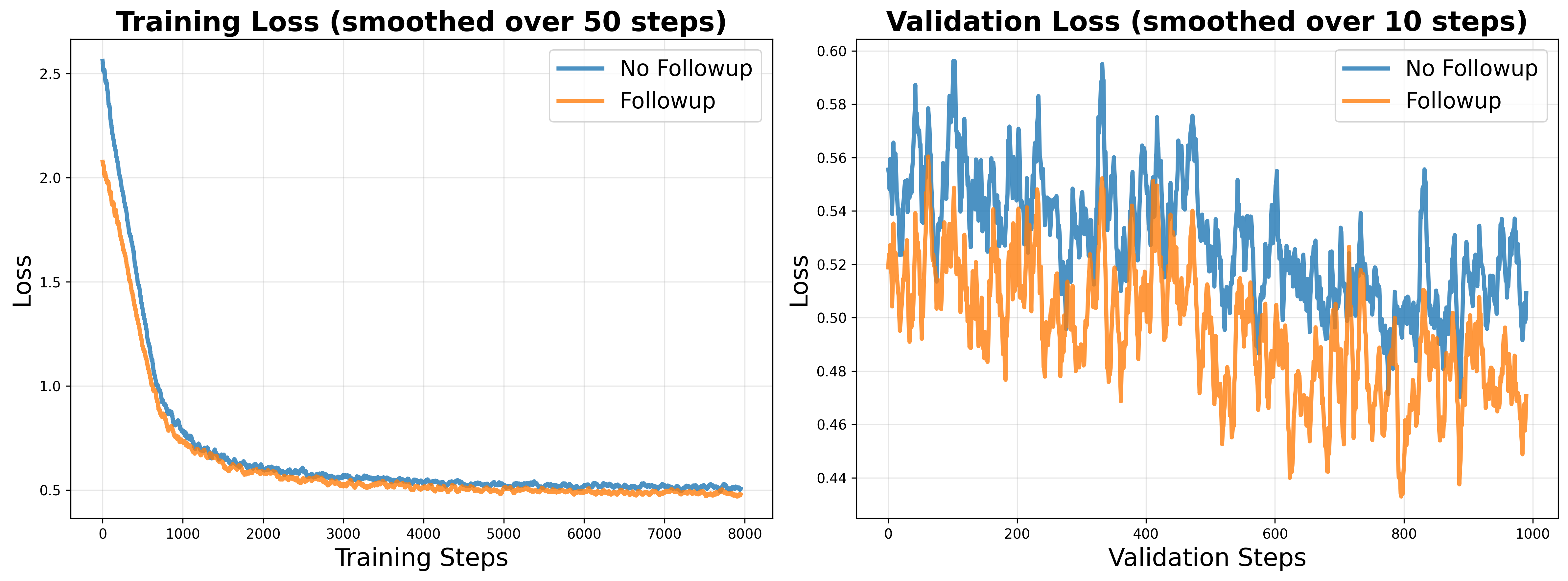}
        \caption{Effect of followup questions on training and validation loss}\label{fig:followup_ablation}
\end{figure}

\begin{figure}[htbp]
    \centering
    \begin{promptbox}[]
        \textbf{General prompt:} \\
        You are an expert Stellar Astrophysicist. You are provided with:
        1. A visual plot of a stellar spectrum.
        2. A raw data summary consisting of two arrays: Wavelength (Angstroms) and Flux (normalized).
        
        Your Goal: precise inference of physical parameters. 
        
        OUTPUT FORMAT: Return ONLY valid JSON.
        \\
            $\{$"spectral type": "string (e.g. G2V)",
            "teff pred": "integer (Kelvin)",
            "logg pred": "float",
            "feh pred": "float (dex)",
            "confidence score": "float (0-1)",
            "key lines detected": ["list", "of", "lines"],
            "reasoning trace": "Brief physics explanation of why you chose these values." $\}$
        \\
        Additional Knowledge Base for Spectral Analysis:
        \textit{$<$Spectra Guidelines$>$}
        Use this knowledge to inform your analysis.
                
        \tcbline
        
        \textbf{Sample question:} \\
        Here are the wavelength and flux data arrays for Star ID 154001001:
        \textit{$<$Wavelength and Flux Values$>$} \\
        Analyze the spectrum.
        
        \tcbline

        \textbf{Gemini-3 pro Answer:} \\
        spectral type: G2V \\
        $T_{eff}$: 5750\\
        $logg$: 4.45\\
        $FeH$: -0.1\\
        confidence score: 0.95\\
        key lines detected: ['Ca II K (3933 A)', 'Ca II H (3968 A)', 'H-delta (4101 A)', 'G-band (4300 A)', 'H-gamma (4340 A)', 'H-beta (4861 A)', 'Mg b (5172 A)', 'Na D (5890 A)', 'H-alpha (6563 A)']\\
        Reasoning trace: The spectrum exhibits the textbook morphology of a solar-type G dwarf. The most prominent features are the deep Ca II H \& K absorption lines (3933/3968 A), which drive the significant break in the continuum around 4000 A. The distinct absorption of the G-band (CH molecule) at 4300 A is a primary classifier for G-type stars. The Hydrogen Balmer series (H-alpha, H-beta, H-gamma) is clearly visible but moderate in strength; they are weaker than in F-types and stronger than in K-types, constraining Teff to approximately 5600-5900 K. The Mg b triplet at 5172 A is visible, and the continuum shape lacks the extreme molecular blanketing of M dwarfs or the blue slope of OB stars. The general 'roughness' of the blue continuum indicates a metallicity consistent with the solar neighborhood ([Fe/H] ~ 0)\\
        \tcbline
        \textbf{True parameters} \\
         spectral type: F5 \\
         $T_{eff}$: 6510\\
         $logg$: 4.17\\
         $FeH$: -0.05\\

    \end{promptbox}
    \caption{An example of the prompt and answer of Gemini-3 pro zero-shot experiment}
    \label{fig:gemini_zero_example}
\end{figure}

\begin{figure}[h]
    \centering
        \centering
        \includegraphics[width=\textwidth]{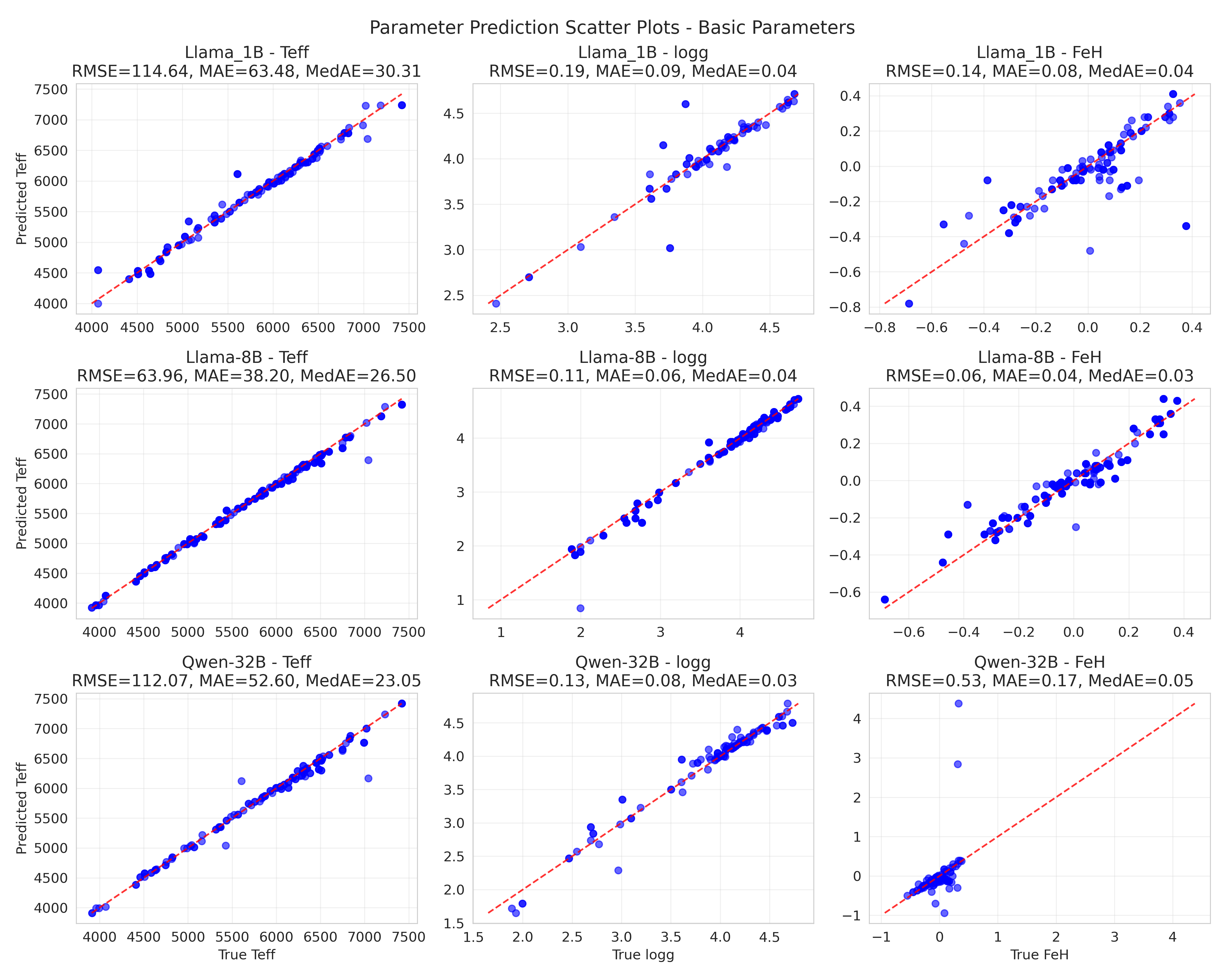}
        \caption{scatter plots of parameter predictions for different models. The columns corresponds to $T_{eff},logg$, and $FeH$. The rows correspond to 1B, 8B, and 32B models.}\label{fig:scatter_basic}
\end{figure}

\begin{figure}[h]
    \centering
        \centering
        \includegraphics[width=\textwidth]{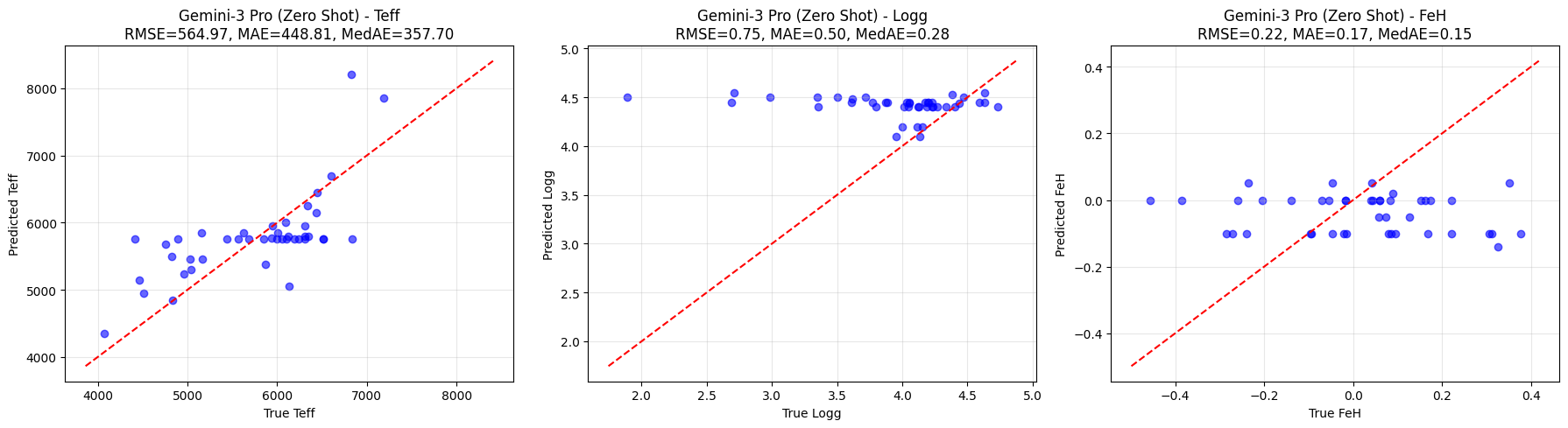}
        \caption{Scatter plots of $T_{eff},logg$, and $FeH$ for Gemini 3 pro without training (zero shot).}\label{fig:scatter_gemini}
\end{figure}

\begin{figure}[ht]
    \centering
    \begin{subfigure}{\textwidth}
        \centering
        \includegraphics[width=0.8\textwidth]{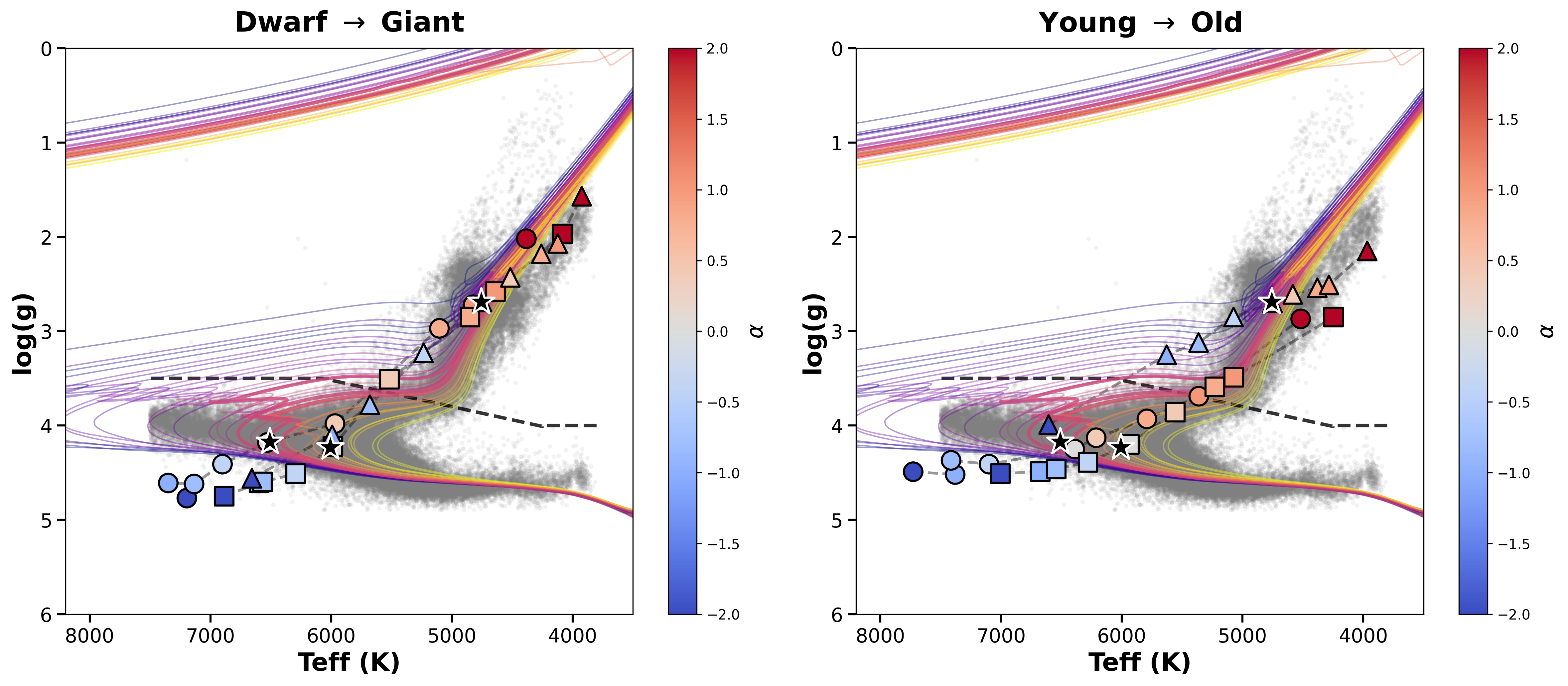}
        \caption{LI-Llama-8B Steering}
        \label{fig:steering_llm}
    \end{subfigure}
    
    \vspace{1em} 

    \begin{subfigure}{\textwidth}
        \centering
        \includegraphics[width=0.8\textwidth]{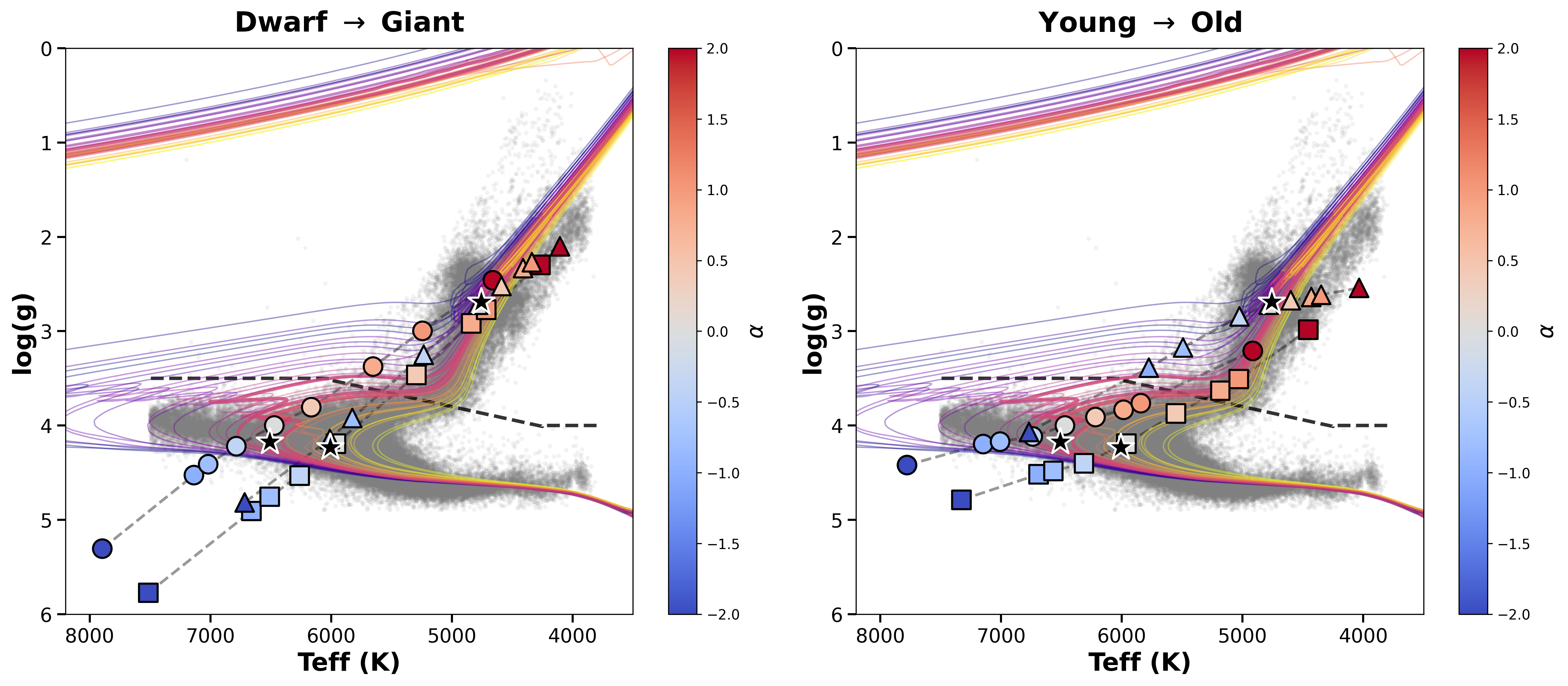}
        \caption{MLP Steering}
        \label{fig:steering_mlp}
    \end{subfigure}

    \caption{Three steering trajectory examples. The upper panel (a) shows our model, and the lower panel (b) shows an MLP trained to predict stellar parameters. Markers (circles, squares, triangles) represent distinct samples. Note that the MLP trajectories frequently enter non-physical regions, whereas our model remains within physically consistent bounds.}
    \label{fig:steering_comparison}
\end{figure}

\begin{figure}[h]
    \centering
        \centering
        \includegraphics[width=\textwidth]{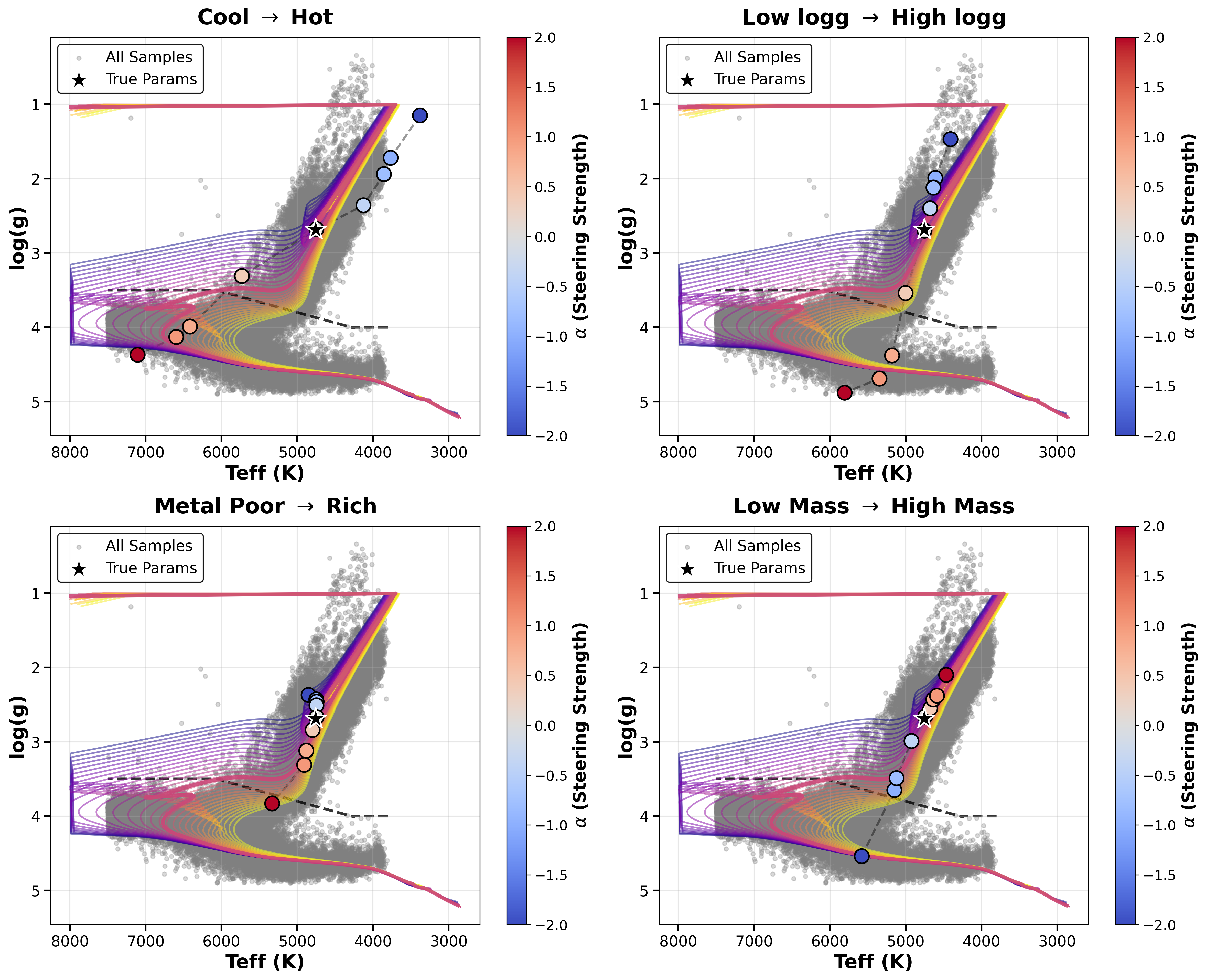}
        \caption{Steering diagrams for the same example as in Figure \ref{fig:steering_example}, and additional concepts.}\label{fig:steering_appendix}
\end{figure}

\end{document}